%% file: main_revtex.tex
\documentclass[%
 aps,
 amsmath,amssymb,
 reprint,%
floatfix,
]{revtex4-1}

\usepackage{amsfonts}
\usepackage{amsmath}
\usepackage{amssymb}
\usepackage{graphicx}
\usepackage{mathrsfs}
\usepackage{color}
\usepackage{bbold}
\usepackage{CJK}
\usepackage{times}
\usepackage{hyperref}
\usepackage{mathtools}

\usepackage{graphicx}
\usepackage{dcolumn}
\usepackage{bm}

\usepackage[utf8]{inputenc}
\usepackage[T1]{fontenc}
\usepackage{mathptmx}
\usepackage{color}

\begin{document}

\title[]{Quasi-ideal feedback-loop supercurrent diode}

\author{Giorgio De Simoni}
\email{giorgio.desimoni@nano.cnr.it}
\affiliation{NEST, Istituto Nanoscienze-CNR and Scuola Normale Superiore, I-56127 Pisa, Italy}
\author{Francesco Giazotto}
\email{francesco.giazotto@nano.cnr.it}
\affiliation{NEST, Istituto Nanoscienze-CNR and Scuola Normale Superiore, I-56127 Pisa, Italy}

\preprint{AIP/123-QED}

\begin{abstract}

\input{abstract}
\end{abstract}

\maketitle

\input{text}

\section*{Acknowledgements}
\input{aknow}



\input{bib}

\end{document}

%% file: abstract.tex

We suggest using a device called the Bootstrap Superconducting Quantum Interference Device (BS-SQUID) to break the reciprocity in charge transport. This device uses magnetic flux back-action to create a nonreciprocal current-voltage characteristic, which results in a supercurrent rectification coefficient of up to approximately 95\%. The BS-SQUID works as a quasi-ideal supercurrent diode (SD) and maintains its efficiency up to about 40\% of its critical temperature. The external magnetic flux can be used to adjust or reverse the rectification polarity. 
Finally, we discuss the finite-voltage operation regime of the SD and present a possible application of our device as a half- and full-wave signal rectifier in the microwave regime.

%% file: text.tex
\section{Introduction}
\label{sec:Introduction}
A charge diode is an electronic device with two terminals that allows an electric current to flow in only one direction while blocking it in the opposite direction, making it a nonreciprocal device.
This property is due to the lack of spatial symmetry, which is intentionally broken during device design and fabrication, \textit{e. g.} in systems such as $pn$ and Schottky junctions. The efficiency of a diode is measured by the saturation current, which determines its degree of ideality through the amplitude of the maximum reverse current when the diode is reverse-biased. These semiconductor devices have a wide range of applications, from detecting and generating photons to signal rectification, and they are essential components of semiconducting electronics.
In addition to semiconducting electronics, superconducting electronics has also undergone significant development due to its superior energy efficiency and larger operation frequency. This development has been made possible by demonstrating many superconducting equivalents of semiconductor devices \cite{Buck1956, McCaughan2014,Likharev1991,Nishino1989,Clark1980,DeSimoni2018,Ritter2021}, including systems that implement non-reciprocal dissipationless Cooper pair transport. 
A supercurrent diode (SD) is a superconducting circuital element that has different amplitudes of positive ($I^+$) and negative ($I^-$) switching critical currents. Similar to the semiconducting case, the ideality of an SD is crucially determined by a rectification coefficient $\eta=\frac{|I^+| - |I^-|}{|I^+| + |I^-|}$ that ranges from 0 (no rectification) to 1 (perfect rectification). Thus, an ideal SD with $\eta=1$ can sustain a supercurrent in one direction and just a \textit{normal} dissipative current in the other.

\begin{figure}[t!]
  \includegraphics[width=0.97\columnwidth]{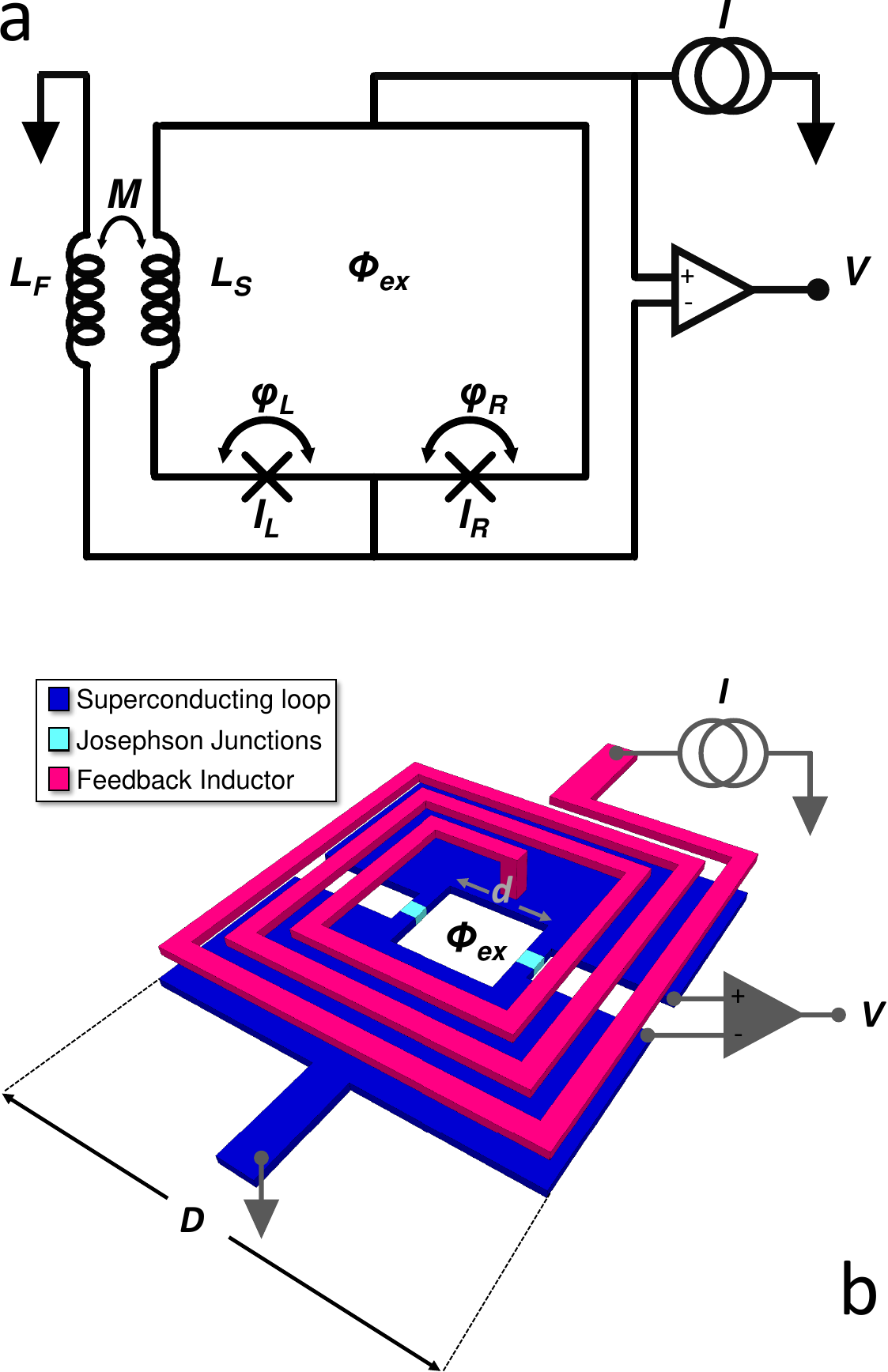}
    \caption{\textbf{Supercurrent diode based on the magnetic-flux back-action: the BS-SQUID} \textbf{a}: Circuital scheme of a supercurrent diode based on the magnetic-flux back-action induced by bootstrapping an external coil. The SQUID is a superconducting ring with total inductance $L_S$, closed on a left and right Josephson junction, with critical current and phase-drop $I_{L}$, $\varphi_L$, and $I_{R}$, $\varphi_R$, respectively. The SQUID is connected in series and inductively coupled via a mutual inductance  $M$ to a feedback inductor, with inductance $L_F$. The series of the SQUID and the inductor realizes the BS-SQUID, which is biased with a current $I$. The output signal of the BS-SQUID ($V$) is measured at the ends of the SQUID. \textbf{b}: Sketch of a possible BS-SQUID implementation: the SQUID (represented in blue) is realized in a \textit{washer} geometry, and is assumed to have $D$ and $d$ as lateral sizes for the whole device and the hole, respectively. The feedback inductance (pink) is assumed to be an electrically insulated squared superconducting spiral overlapping the SQUID. The Josephson junctions are represented in light blue.}
  \label{fig:fig1}
\end{figure}
Although the idea of realizing supercurrent rectifiers is in itself not new \cite{10.1063/1.4790625} and despite a recently renewed \cite{ando2020observation} and intense research effort, a platform of choice to realize SDs has not been identified yet. Indeed, nonreciprocal supercurrent transport, due to inversion and time-reversal symmetry breaking, was demonstrated in relatively exotic materials and heterostructures \cite{wakatsuki2017nonreciprocal,diez2023symmetry,lin2022zero,ando2020observation,pal2022josephson,bauriedl2022supercurrent,baumgartner2022supercurrent,jeon2022zero,sundaresh2023diamagnetic,turini2022josephson,daido2022intrinsic}. Several mechanisms, mainly exploiting spin-orbit coupling and magnetic effects, have been proposed to implement rectification in superconducting films and Josephson junctions (JJs) \cite{nadeem2023,daido2022intrinsic,ilic2022theory,yuan2022supercurrent,he2022phenomenological,misaki2021theory,davydova2022universal,zhang2022general,scammell2022theory,wambaugh1999superconducting,vodolazov2005superconducting,villegas2005experimental,van2005vortex,cerbu2013vortex,golod2021reconfigurable,lyu2021superconducting,golod2022demonstration,suri2022non,chahid2023high,hou2023ubiquitous,satchell2023supercurrent,vavra2013current,Margineda2023,arXiv:2311.14503v2}. Nonetheless, the race to maximize the supercurrent rectification efficiency has driven the demonstration of devices with $\eta$ values in the range of a few tens percent \cite{hou2023ubiquitous}, and barely approaching 90\% in Nb devices exploiting current back-action mechanisms \cite{arXiv:2311.14503v2}. Present technology SDs are, therefore, far from ideal. This severe performance limitation, together with the use of complex material systems and the exploitation of physical mechanisms that are hardly modeled and engineered, is holding back the application of SDs in practical applications, limiting the design of new platforms based on their exploitation.

Here, we propose and analyze a superconducting quantum interference device (SQUID) that utilizes a feedback loop to implement a magnetic flux back-action. This results in a current-voltage characteristics of the device that strongly favors one direction over the other. By adjusting the external magnetic flux $\Phi_{ex}$, we can unbalance the critical currents $I^+$ and $I^-$ to achieve a nearly perfect rectification coefficient, which allows us to implement a quasi-ideal SD.

\section{Bootstrap SQUID supercurrent diode}
\label{sec:Bootstrap_SQUID_supercurrent_diode}
In their more conventional implementation, which we assume in the following, SQUIDs consist of two tunnel JJs closed on a superconducting loop, whose switching current is modulated by the magnetic flux $\Phi$ threading the device with a periodicity equal to the magnetic flux quantum, $\phi_0 = h/2e$ \cite{Clarke2004}. Thanks to their excellent sensitivity, SQUIDs are at the core of state-of-the-art cryogenic magnetometers, inductively-coupled current amplifiers \cite{Clarke2004,Barone1982,Kleiner2004,Martinez-Perez2017,Granata2016,Fagaly2015}, and are often included in larger systems designed to perform signal processing applications \cite{Kornev2017a}. The use of SQUIDs as supercurrent rectifiers has been recently shown in systems with a large inductance \cite{paolucci2023gate,10.1063/5.0165259}, which on one side is crucial to produce the rectifying behavior, while on the other side limits the maximum achievable rectification, due to the reduced amplitude of the switching current modulation with the flux \cite{Clarke2004}. 
Our approach relies on coupling the SQUID loop with a feedback coil. Feedback inductors are conventionally bootstrapped in the control and readout circuitry of SQUID amplifiers and magnetometers to reduce the noise contribution from the room-temperature preamplifiers and increase their linear dynamic range. \cite{10.1063/1.103650,MUCK2002141,133723,10.1063/1.114499,5ecc143f09064b7ca70c9bd700bb9f42,0fc1f52cc62447a3b25e1f946ca70262,Xie_2010,WANG2012127,Zhang_2013,PhysRevApplied.19.054021,PhysRevApplied.18.014073}. Similarly, in our approach, a feedback coil with inductance $L_F$ is wired in series and inductively coupled via a mutual inductance coefficient $M$ to a current-biased SQUID, consisting of a ring of inductance $L_S$ and a left and a right JJs having critical currents and phase drops $I_L$ and $I_R$ and $\varphi_L$ and $\varphi_R$, respectively. The total magnetic flux $\Phi$ threading the loop is the sum of the external flux $\Phi_{ex}$ and of the flux induced by the feedback loop $\Phi_F=MI$, where $I$ is the bias current flowing through the interferometer. It follows that $\Phi$ is a function of $I$, which is common to both the SQUID and the feedback coil. This results in a flux back-action mechanism, which can be exploited to implement the supercurrent rectifying mechanism. The circuital scheme of such a bootstrap SQUID (BS-SQUID) is depicted in Fig. \ref{fig:fig1}a.

To understand where the rectification properties arise from, it is necessary to resort to the resistively shunted Josephson junction (RSJ) equations of the SQUID \cite{Clarke2004}:
\begin{equation}
\label{equation:RCSJ}
\begin{split}
\frac{I}{I_0}=(1-\alpha_i)sin(\varphi_L)+(1+\alpha_i)sin(\varphi_R),
\\
\frac{2j}{I_0}=(1-\alpha_i)sin(\varphi_L)-(1+\alpha_i)sin(\varphi_R),
\\
\delta \varphi=\varphi_L-\varphi_R=2 \pi \Phi+\pi \beta_l(j-\alpha_l I),
\\
I_0=\frac{I_L+I_R}{2},
\\
L_S=L_L+L_R.
\end{split}
\end{equation}
In Eqs. \ref{equation:RCSJ}, we introduced the coefficients $\alpha_i=\frac{I_R-I_L}{I_R+I_L}$, $\alpha_l=\frac{L_R-L_L}{L_R+L_L}$, and $\beta_l=\frac{2 L_S I_0}{\Phi_0}$   accounting for the difference between the critical current of the left and right junction,  the asymmetries in the inductance ($L_L$ and $L_R$) of the two arms of the loop, and the SQUID screening, respectively. $I^+$ and $I^-$ are calculated from Eqs. \ref{equation:RCSJ} by maximizing and minimizing $I$ over $\delta \varphi$, respectively.
By expanding the fluxoid quantization relation
\begin{equation}
\label{equation:mod_flux_quant}
\begin{split}
\varphi_L-\varphi_R=2 \pi (\Phi_{ex}+MI)+\pi \beta_l(j-\alpha_l I)
\end{split}
\end{equation}
to account for the flux contribution from the feedback inductor, we observe that the latter is mathematically equivalent to unbalancing the inductance of the SQUID arms. This is known to result in a skewing of the positive $I^+(\Phi_{ex})$  and negative $I^-(\Phi_{ex})$ critical current \textit{vs.} flux characteristics  in opposite magnetic flux directions, thereby causing a nonreciprocal behavior and, at fixed external flux, a supercurrent diode effect. We note, that the introduction of an asymmetry in the SQUID arms inductances can be exploited to implement or tune a SD \cite{paolucci2023gate}. Nonetheless, it is worth emphasizing that $|\alpha_l|$ is by definition bound to 1, and that the $I^{+,-}(\Phi_{ex})$ skewness can be enhanced by increasing the total SQUID inductance, but at the cost of damping the amplitude of the $I^{+,-}(\Phi_{ex})$ swing. It is possible to achieve a rectifying behavior by using the construction parameter of the SQUID inductance, but this behavior is modest. On the other hand, introducing the feedback coil creates an additional capability to control the mutual inductance coefficient, denoted as $M$, which is still proportional to $L_S$ and can be adjusted through the value of $L_F$. This provides the opportunity to modify the skewness of the BS-SQUID $I^{+,-}(\Phi_{ex})$ without significantly affecting the amplitude of their oscillation.

To illustrate this statement, we assume that our BS-SQUID has a \textit{washer} geometry, shown in Fig. \ref{fig:fig1}b, with the feedback inductor, comprising a spiral with $n$ coils, overlapped onto the SQUID. It can be shown \cite{10.1063/1.93210,1060902} that, by neglecting the parasitic inductance associated with the Josephson junctions it holds
\begin{equation}
\label{equation:washer}
\begin{split}
L_S\simeq 1.25\mu_0 d,
\\
M\simeq n L_S=\frac{n \beta _l \Phi_0}{2 I_0},
\end{split}
\end{equation}
where $\mu_0$ is the vacuum magnetic permeability and $d$ the lateral size of the SQUID  hole. It is worth noting that, with this geometric choice, $M$ is proportional to the product $n\times L_S$ and ($n\times \beta_l$). The number of coils that can be overlapped to the SQUID is a function of the washer lateral size $D$, while $L_S$ is proportional only to its inner size $d$. This results in the possibility of increasing the mutual inductance coefficient without increasing the SQUID inductance. In turn, this allows to increase the skewness of the $I^{+,-}(\Phi_{ex})$ of the BS-SQUID without significantly affecting the critical current modulation amplitude.
\begin{figure}[t!]
  \includegraphics[width=1.0\columnwidth]{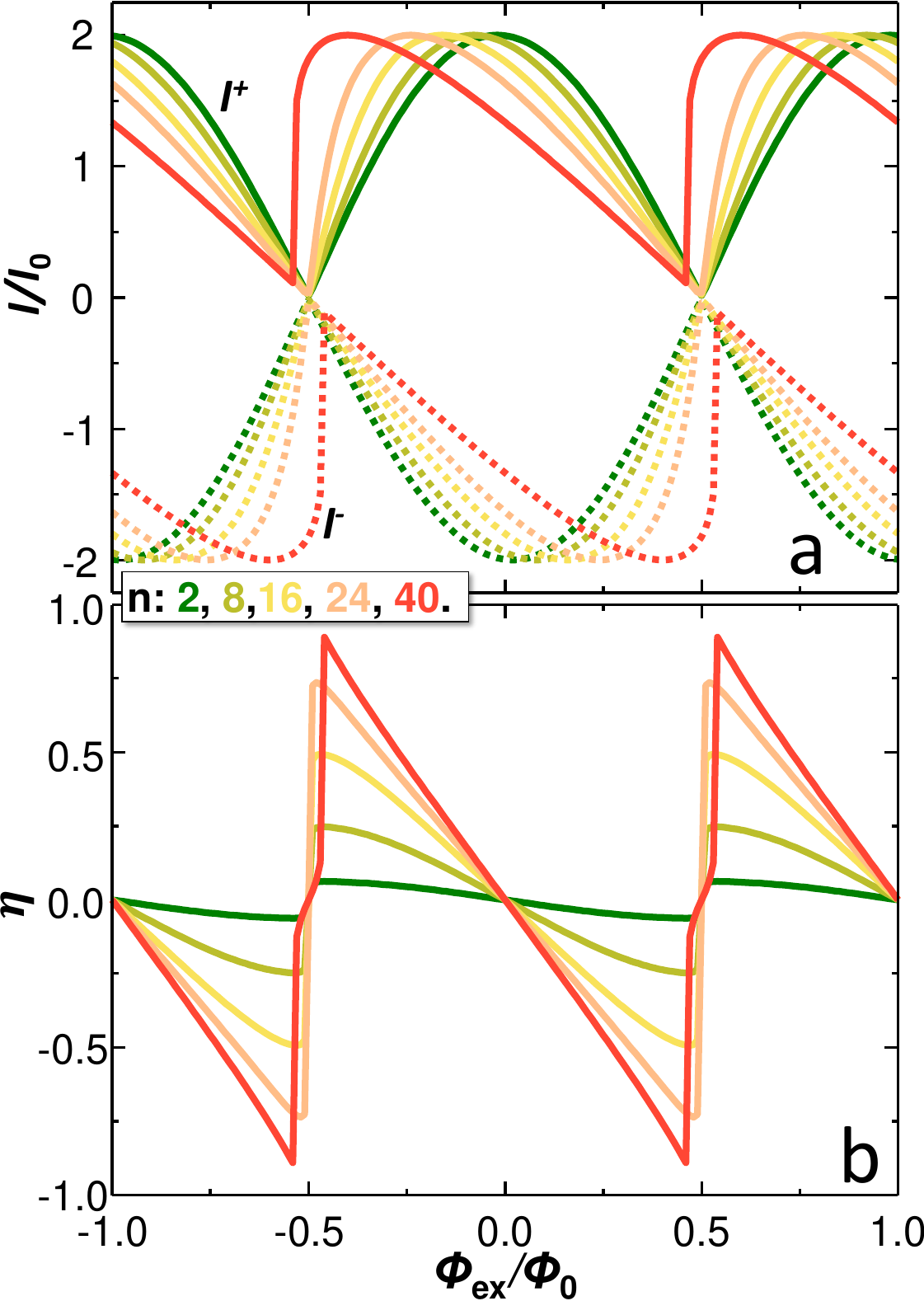}
    \caption{\textbf{BS-SQUID as supercurrent diode} \textbf{a}: Positive $I^+$ (solid curves) and negative ($I^-$, dashed curves) normalized switching current of a \textit{washer} BS-SQUID \textit{vs.}  external magnetic flux $\Phi_{ex}$ for selected values of number of windings $n$ of the the feedback inductor. The BS-SQUID is assumed to have identical Josephson junctions ($\alpha_i=0$) and symmetrical inductance of the arms ($\alpha_l$=0). The screening parameter $\beta_l$ is set to 0.01. \textbf{b}: Rectification coefficient $\eta$ \textit{vs.} $\Phi_{ex}$ calculated from the switching current characteristics of panel a.}
  \label{fig:fig2}
\end{figure}

\section{Rectification performance}
\label{sec:Rectification performance}
Figure \ref{fig:fig2}a shows the normalized values $I^+$ and $I^-$ \textit{vs.} $\Phi_{ex}$ for selected values of $n$ between 2 and 40, when $\beta_l=0.01$ and $\alpha_i=\alpha_l=0$. Due to the low value of the screening parameter, at low winding numbers of the feedback coil (\textit{e. g.} dark green curve), the $I^{+,-}(\Phi_{ex})$ is almost equivalent to that of a perfectly symmetric SQUID, with a negligible skewing and with an almost full-span ($\sim 2$) modulation amplitude (the latter is limited only by  $I^+$ and $I^-$ never being exactly 0). By increasing $n$, the maxima of $|I^+|$ ($|I^-|$) are progressively pulled toward negative (positive) values of the magnetic flux while the minima (maxima) remain almost locked in the original position. These characteristics yield the aforementioned skewing of the $I^{+,-}(\Phi_{ex})$, which is the key to achieving an efficient SD effect, and are present up to $n\sim40$. This results in a strong and tunable supercurrent rectifying behavior, whose amplitude and sign can be controlled through the external magnetic flux. 
This can be well appreciated by plotting $\eta$ \textit{vs.} $\Phi_{ex}$ for the same selected values of $n$ (see Fig. \ref{fig:fig2}b): the BS-SQUID exhibits a periodic modulation of the rectification coefficient, which changes its sign at $\Phi_{ex}=N \frac{\Phi_0}{2}$, with $N=0,\pm 1,\pm 2...$. 
This peculiar feature, which has been reported so far in just a few cases \cite{pal2022josephson,sundaresh2023diamagnetic,arXiv:2303.01902,costa2023,kawarazaki2022magnetic,gupta2023gate,paolucci2023gate}, can be exploited to reverse the polarity of the device during its operation. Furthermore, by increasing $n$ the maximum value of $\eta$ increases as well, due to the maxima of $|I^+|$ and $|I^-|$ being pulled in different magnetic flux directions. For $n\lesssim 40$, $\eta$ reaches its maximum value, which is just slightly lower than 1.

It is impossible to achieve ideal rectification due to the requirement of a low, but non-zero, value of the SQUID inductance. However, we believe that if we can optimize the inductances or implement the device concept differently, we might be able to improve the rectification performance beyond what is shown in this work and closer to the ideality limit. In our geometry, when $n\gtrsim 40$, the excessive distortion of the $I^{+,-}(\Phi_{ex})$ causes $\eta$ to stop increasing. This is mainly because the minima of $|I^+|$ and $|I^-|$ are finally unblocked and raising, as shown by the red curve in Fig. \ref{fig:fig2}.

\begin{figure}[t!]
  \includegraphics[width=0.97\columnwidth]{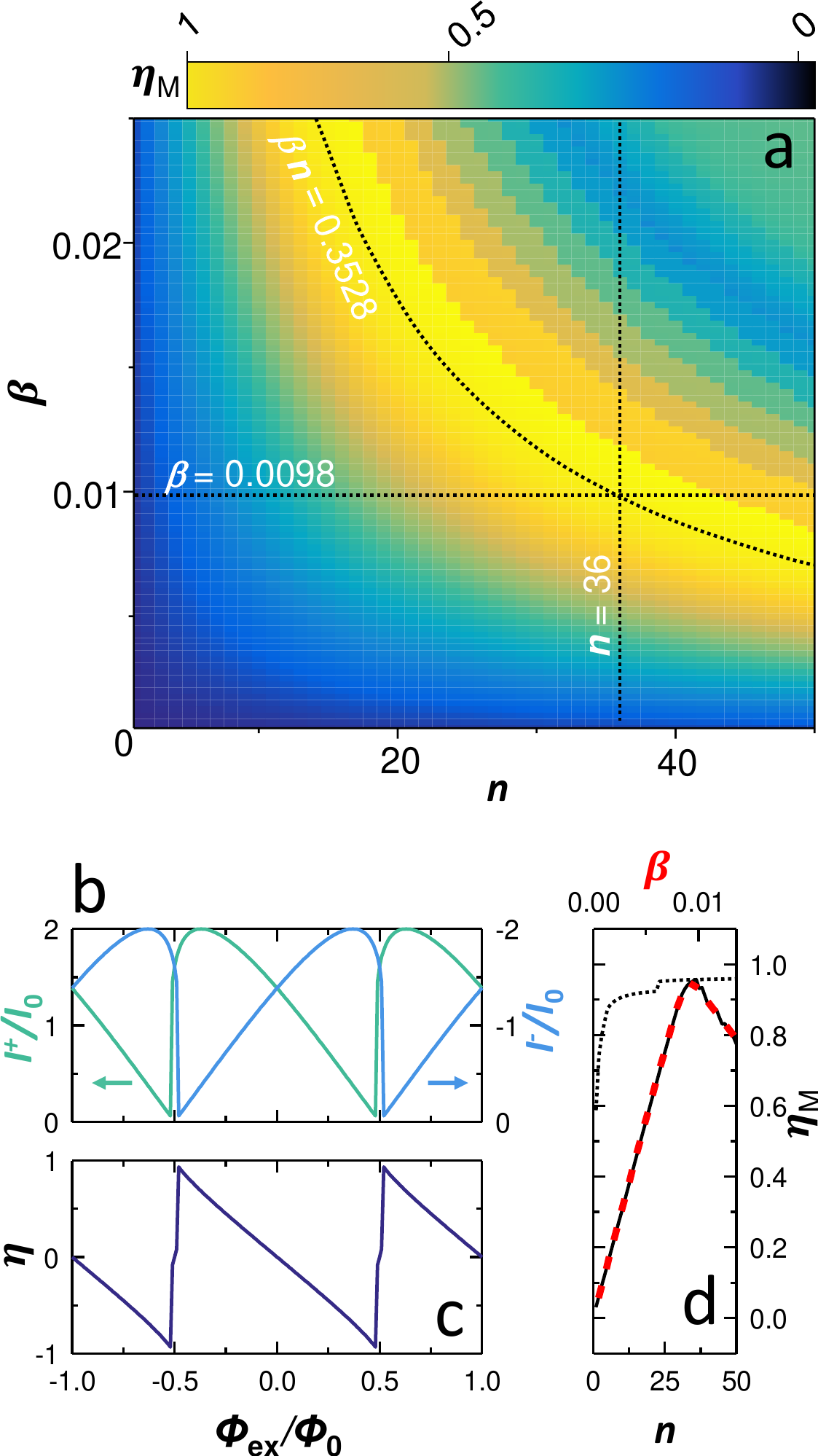}
    \caption{\textbf{Characterization of maximum rectification}. \textbf{a}: Maximum value of the rectification $\eta_M$ \textit{vs.}  winding number of the feedback inductance $n$ and screening parameter $\beta_l$ of a perfectly symmetrical BS-SQUID. 
    Dashed black curves highlight the cut-lines at constant $\beta_l=0.098$, $n=36$, and $\beta_l n=0.3526$. Where such cut-lines cross maximum rectification is achieved; in particular, along the latter  $\eta_m$ remains almost constant. \textbf{b}: Plot of $I^+$ (left scale) and $|I^-|$ (right scale) \textit{vs.} $\Phi_{ex}$, calculated at $n=36$ and $\beta_l=0.0098$, corresponding to the cross point of the cut-lines of panel b. \textbf{c}: $\eta$ \textit{vs.} $\Phi_{ex}$ calculated for the same parameter choice of panel b, \textit{i. e.} where maximum and minimum supercurrent rectification are achieved (at $\Phi_{ex}\sim \pm0.5\Phi_0$). \textbf{d}: Plot of $\eta_m$ along the cut-lines of panel a, \textit{vs.} $\beta_l$ (dashed red line, top scale) and $n$ (short-dotted and continuous black lines, bottom scale). The red dashed line is the plot of $\eta_M$ calculated along the $n=36$ cut-line. The short-dotted line is the plot of $\eta_M$ calculated along the $\beta_l n$=0.3528 cut-lines. The continuous black line is the plot of $\eta_M$ calculated along the $\beta_l=0.0098$ cut-line.}
  \label{fig:fig3}
\end{figure}

In Fig. \ref{fig:fig3}a, we plot the rectification efficiency maximum value $\eta_M$ \textit{vs.} $n$ and $\beta_l$. The hyperbolic profile of the $\eta_M$ intensity levels shows how it is essentially controlled by the mutual induction parameter $M$: $\eta_M$ grows both as $n$ and $\beta_l$ increase until it reaches the maximum value of approximately $90 \%$, when $\beta n \sim 0.35$. 
Above such a threshold $\eta_M$ decreases due to the excessive distortion of the $I^{+,-}(\Phi_{ex})$s. Figures \ref{fig:fig3}b and c show $I^{+,-}$ and $\eta$ \textit{vs.} $\Phi_{ex}$, respectively, calculated by setting $\beta_l=0.0098$ and $n=36$ in our model. In this spot, $\eta_M$ reached its maximum ( $\sim 0.96$), a value truly close to the ideality limit and among the highest reported so far. Based on this remarkable result and their constructive simplicity, we believe that BS-SQUIDs are excellent candidates for the implementation of superconducting platforms requiring supercurrent rectification. Interestingly, the evolution of $\eta_M$ due a variation of to either $\beta_l$ or $n$ follows a similar and almost identical functional form (see Fig. \ref{fig:fig3}d, solid-black and red-dashed lines), thereby confirming $M$ as the leading parameter in determining the rectification performance. Although changing $\beta_l$ ($n$) at constant $n$ ($\beta_l$) results in a fast decrease of the rectification coefficient, it is instead almost constant if the product $\beta_l \times n$ is kept constant, with $25 \lesssim n \lesssim 50$. When $n\lesssim 25$, the rectification drops again due to a low magnetic coupling and the resulting scarce back-action (see Fig. \ref{fig:fig3}d, dotted-black line). 

\section{Impact of temperature and device imperfections on the rectification efficiency}
\label{sec:temp_and_alpha}
In the following we discuss the impact of the temperature and of device imperfections, due to fabrication limits, on the rectification efficiency of the BS-SQUIDs.

\begin{figure}[t!]
  \includegraphics[width=1\columnwidth]{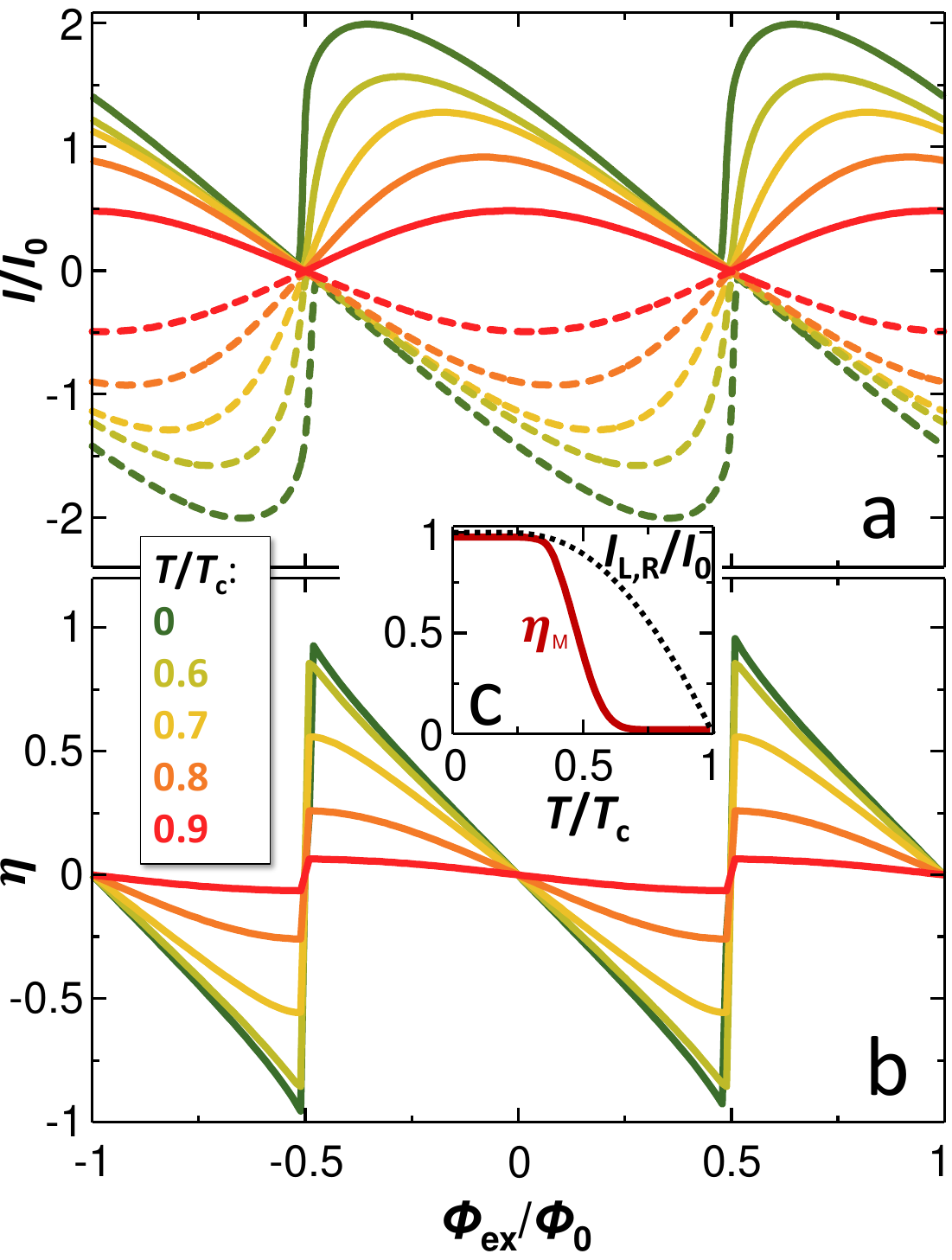}
    \caption{\textbf{Impact of the temperature.} \textbf{a}: $I^+$ (solid curves) and $I^-$ (dashed curves) \textit{vs.} $\Phi_{ex}$ for selected values of the temperature $T$, between 0 and 0.9 $T_C$, calculated for a perfectly symmetrical BS-SQUID with $\beta_l=0.0098$ and $n=36$. \textbf{b}: Rectification coefficient $\eta$ \textit{vs.} $\Phi_{ex}$ for same values of $T$ selected for panel a. \textbf{c}: Maximum rectification coefficient  $\eta_M$ (solid red curve) and Josephson critical current $I_{L,R}$ (dashed black line) \textit{vs.} temperature $T$.}
  \label{fig:fig4}
\end{figure}
To evaluate the resilience of the supercurrent rectification capability of BS-SQUIDs to the temperature it is necessary to take into account the temperature dependence of the critical current of the Josephson junctions. The latter is described by the Ambgaokar-Baratoff relation (AB) \cite{AB1963,AB1963erratum,tinkham2004introduction}, 
\begin{equation}
    \begin{split}
          \frac{I_{L,R} (T)}{I_0}= \frac{\Delta (T)}{\Delta(T=0)} tanh \left( \frac{\Delta(T)}{ 2 k_B T} \right),
    \end{split}
\end{equation}
where $I_0$ is the zero-temperature critical current, $\Delta(T)/\Delta(T=0)$ is the universal normalized Bardeen-Cooper-Schrieffer temperature-dependent amplitude of the superconducting energy gap \cite{tinkham2004introduction} of the superconducting SQUID loop, and $k_B$ is the Boltzmann constant. Although in our model we take into account the SQUID inductance (and mutual inductance) through the screening parameter $\beta_l$, to compute the temperature response of the BS-SQUID we decided to keep constant the inductance of the loop ($L_S=\beta_{l0} \Phi_0 / 2 I_0$, with $\beta_{l0}=0.0098$) and the mutual inductance ($M=nL_S=36\times L_S$). In the case in which the kinetic inductance of the SQUID can be neglected, this choice provides more physical insight, since $L_S$ and $M$ are determined by geometric parameters, which remain constant in  temperature, while $\beta_l$ varies due to its dependence on $I_{L,R}(T)$. 

Figure \ref{fig:fig4}a shows the positive and the negative switching currents of the BS-SQUID vs. external magnetic flux at selected values of $T$. Following from the temperature evolution of $I_{L,R}(T)$ (see Fig. \ref{fig:fig4}c, black-dashed curve), the amplitude of the switching current modulation remains constant up to $\sim0.4 T_C$, above this threshold it decreases to vanish at $T_C$. In particular, the modulation amplitude is about one-half of its zero-temperature value at $T=0.7 T_C$, and $\sim\frac{1}{4}$ at $T=0.9 T_C$. Through the impact of $I_C(T)$ on $\beta_l$, the skewness of the curves is also almost preserved up to the same temperature, above which it is progressively reduced resulting in a symmetrization of the positive and negative characteristics. This is directly reflected in the rectifying behavior of the device. In Fig. \ref{fig:fig4}b we plot $\eta$ \textit{vs} $\Phi_{ex}$ for the same values of $T$ selected for Fig. \ref{fig:fig4}a. Although the pattern of $\eta$ remains essentially unchanged as the temperature rises, its amplitude decreases accordingly with the evolution of the $I^{+,-}(\Phi_{ex})$ characteristics. It is worth noting that, due to the double impact of the reduction of $I_{L,R}$ on both the maxima of $|I_S^{+,-}|$ and on $\beta_l$, the reduction of $\eta_M$ with the temperature is much faster than that of $I_{L,R}$, resulting in $\eta_M\sim0.5\eta_M(T=0)$ at $T=0.7T_C$, but $\eta_M\sim0.25\eta_M(T=0)$ already at $T=0.8T_C$. This is well appreciated through the comparison of $\eta_M(T)$ (red-solid curve) and $I_{L,R}(T)$ shown in Fig. \ref{fig:fig4}c. On this point, it is relevant to emphasize that this characteristic results in the interesting feature to obtain an almost constant rectifying behavior up to $T\sim 0.4 T_C$, \textit{i. e.} where $I_{L,R}(T)$ is only marginally affected by heating. Furthermore, the supercurrent rectification is  exploitable up to a temperature very close to the critical one, with $\eta_M$ being still about $5\%$ at $T\sim0.9 T_C$.

\begin{figure}[t!]
  \includegraphics[width=0.97\columnwidth]{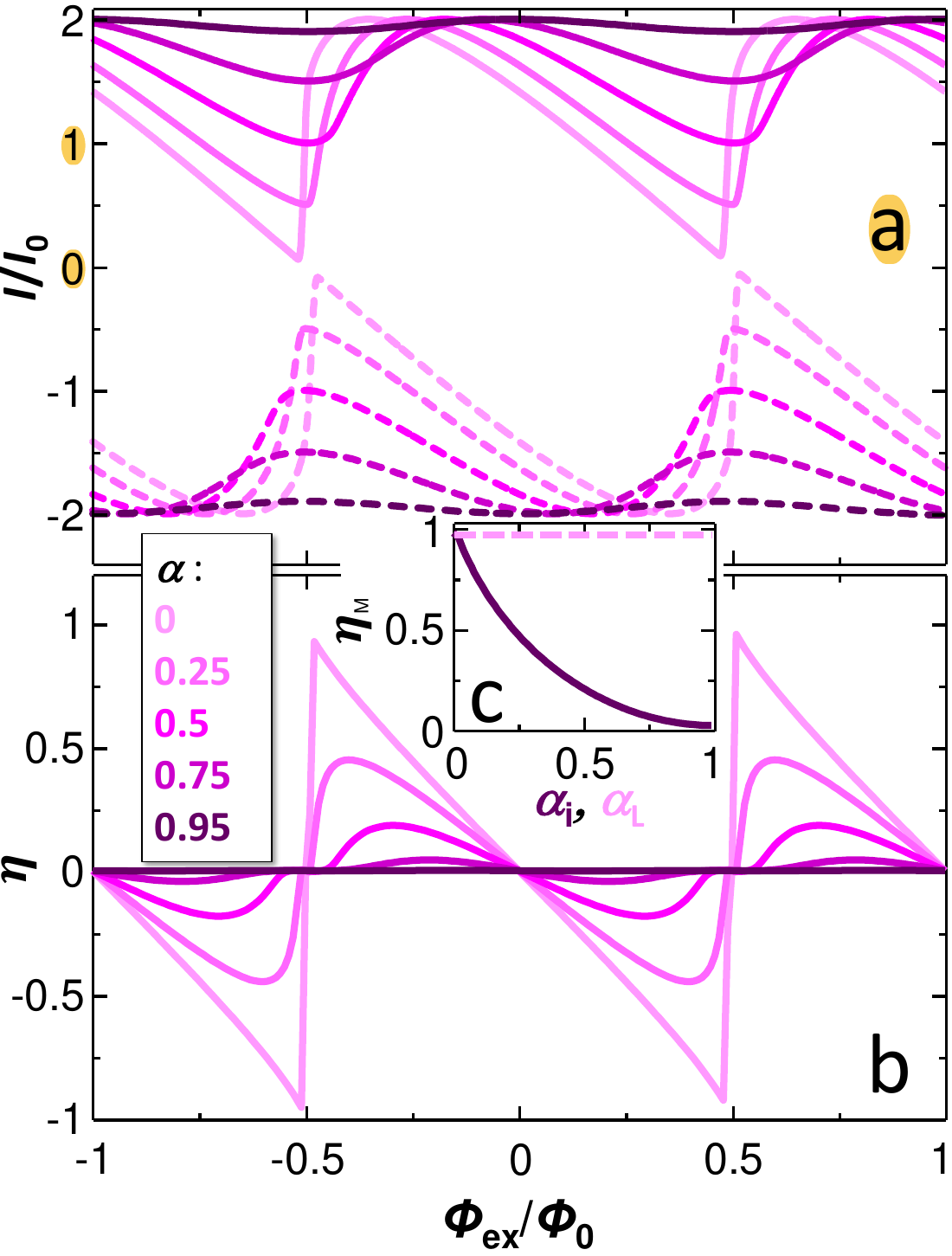}
    \caption{\textbf{Impact of device imperfection.} \textbf{a}: $I^+$ (solid curve) and $I^-$ (dashed curve)\textit{vs.} $\Phi_{ex}$ for selected values of  junction critical current asymmetry $\alpha_i$, between 0 and 0.95, calculated with $\beta_l=0.0098$ and $n=36$. \textbf{b}: Rectification coefficient $\eta$ \textit{vs.} $\Phi_{ex}$ calculated for the same values of $\alpha_i$ selected for panel a. \textbf{c}: Maximum rectification efficiency $\eta_M$ \textit{vs.}  $\alpha_i$ (solid violet curve) and asymmetry $\alpha_l$ of the inductance of SQUID arms (dashed pink curve).}
  \label{fig:figalpha}
\end{figure}

Our model includes two parameters, $\alpha_i$ and $\alpha_l$, to account for imperfections in the device. These imperfections arise from asymmetries in the critical current of the left and right JJ, which can be caused by differences in junction area or tunnel barrier resistivity. Additionally, an unbalanced inductance between the two arms of the SQUID can also impact the device. However, we will discuss that this does not significantly affect the rectification properties of the BS-SQUID. It is important to note that having good symmetry between the JJs is a major requirement for optimal device performance.
Figure \ref{fig:figalpha}a, shows the modification of the $I^{+,-}(\Phi_{ex})$ characteristics, calculated at $T=0$ for $\beta_l=0.0098$ and $n=36$, when $\alpha$ is raised from 0 to 0.95. As conventionally observed in SQUIDs, a sizable difference between the critical currents of the JJs results in a suppression of the ability of the device to interfere and in the following suppression of the visibility of the modulation of $I_S^{+,-}$ with the flux. This translates into an increasingly smoother $\eta(\Phi_{ex})$ relation as $\alpha_i$ is raised (see Fig. \ref{fig:figalpha}b). 
There are two sides to the mechanism being discussed. On one hand, it can be used positively by introducing a slight asymmetry between the junctions to widen the rectification sweet spot and reduce the device's sensitivity to flux noise. On the other hand, this comes at the cost of a severe drop in the rectification coefficient, as shown in figure \ref{fig:figalpha}c (violet-solid curve). Therefore, it is necessary to have excellent control over the symmetry of the BS-SQUID junctions when designing and creating these devices. This is the most important parameter that needs to be taken care of.

It has been previously mentioned that $\alpha_L$ has no significant impact on the device properties. This has been confirmed by the plot of $\eta_M(\alpha_L)$ shown in Fig. \ref{fig:figalpha}c (dashed-pink line). The reason for this is that an asymmetry in the inductance of the two arms of the SQUID loop has the same mathematical effect as a variation of the mutual inductance. With the values selected for $n$ and $\beta_l$, the variation of the mutual inductance is negligible. Therefore, even at high inductance asymmetries, there is only a slight modification of the magnetic backaction, which is not perceptible. This makes the device extremely robust to fluctuations in this parameter.

\section{Dissipative operation}
\label{sec:dissipative operation}
Practical DC-SQUIDs are conventionally operated in a dissipative fashion, as they are usually biased close to or above their critical current. That is, the SQUID is driven in a dissipative or finite-voltage state. In the following, we wish to discuss the response of BS-SQUID in such a regime. In this configuration, the SQUID response can be analytically calculated following the relation \cite{Clarke2004}
\begin{equation}
\label{eq:McCumber}
\begin{split}
\frac{V}{V_0}=\sqrt{I^2-4I_0^2},
\end{split}
\end{equation}
where $V$ is the voltage drop across the BS-SQUID (see the biasing scheme in Fig. \ref{fig:fig1}a), $V_0=\frac{R_T}{2I_O}$ is a critical voltage, calculated through the tunnel resistance $R_T$ of the JJs. Equation \ref{eq:McCumber} holds in the limit $\frac{2 \pi }{\Phi_0} I_0 R_T^2 C \ll 1$ \cite{Clarke2004,mccumber1968effect}, where $C$ is the capacitance of each JJ. This condition can be usually obtained through the adoption of shunt resistors with resistance $R$ such that $R < R_T$. In the latter case, $V_0$ has to be accordingly modified to account for the parallel of the shunt resistor and of the JJ.

\begin{figure}[t!]
  \includegraphics[width=1\columnwidth]{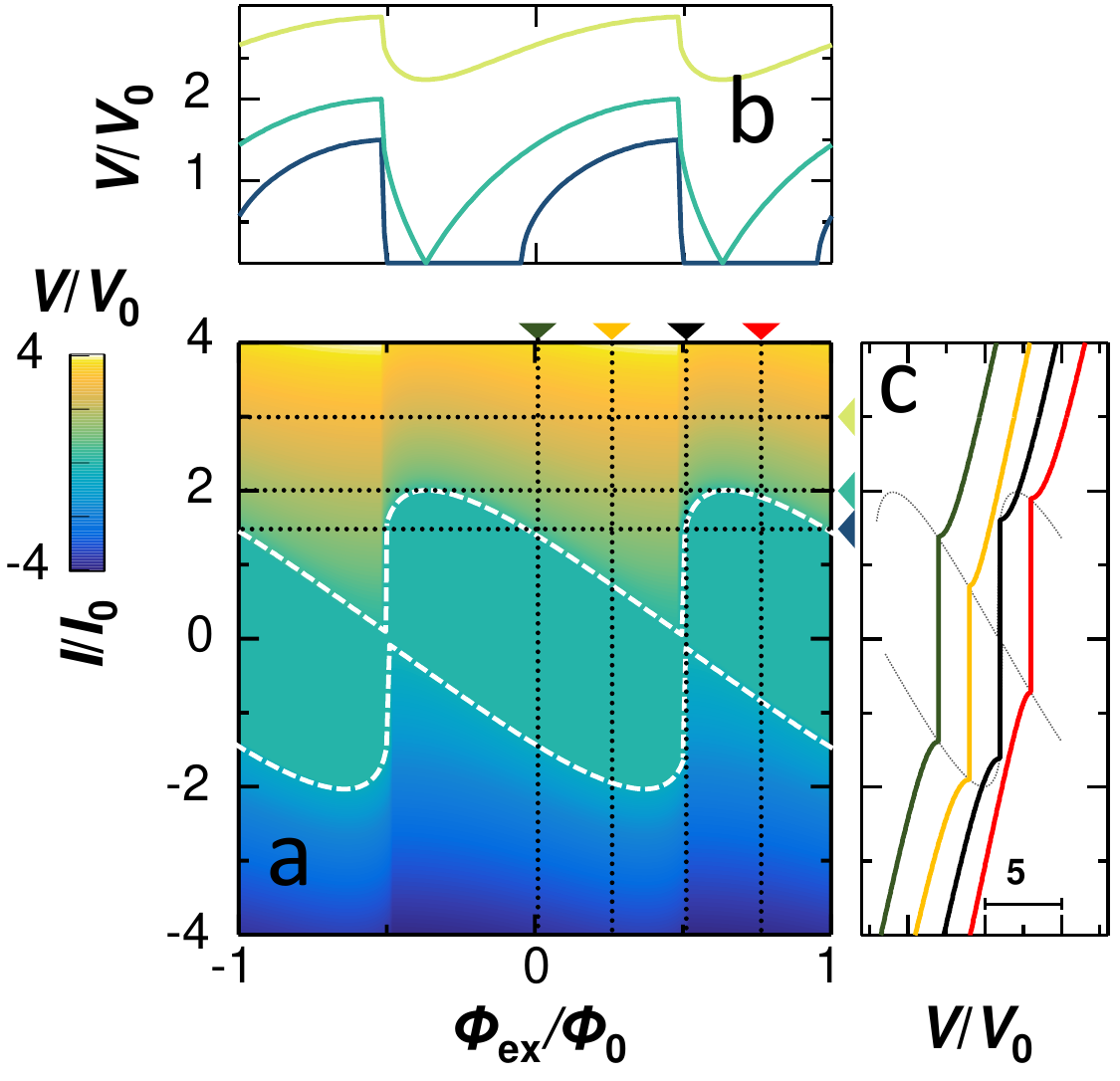}
    \caption{\textbf{BS-SQUID response in the dissipative regime.} \textbf{a}: Normalized voltage drop of a BS-SQUID \textit{vs.} $\Phi_{ex}$ and  bias current $I$, calculated for a perfectly symmetric device with $\beta_l=0.0098$ and $n=36$. The superconducting phase can be identified by the 0-voltage region colored in aquamarine and bounded by the white dashed lines. Dashed black lines correspond to the curves at constant current or constant flux, plotted in panels b and c, respectively. The color of the markers at the edge of the color plot identifies the curve in panels b and c with the same colors. \textbf{b}: $V$ \textit{vs.} $\Phi_{ex}$ curves for selected values of $I$ corresponding to the cut-lines at constant current of panel a. \textbf{c}: $V$ \textit{vs.} $I$ curves for selected values of $\Phi_{ex}$, corresponding to the cut-lines at constant flux of panel a. For the sake of clarity, the curves are horizontally offset by a quantity proportional to the correspondent flux. The dashed gray lines are a guide for the eye, which follow the evolution of $I^+$ and $I^-$ with $\Phi_{ex}$.}
  \label{fig:fig5}
\end{figure}
Figure \ref{fig:fig5}a shows in a color plot the voltage response of a BS-SQUID \textit{vs.} $\Phi_{ex}$ and $I$, when $n=36$ and $\beta_l=0.098$. The aquamarine area corresponds to the non-dissipative regime. When the bias current exceeds the positive (yellowish area) or the negative (blueish area) switching current of the BS-SQUID a finite-voltage response is obtained. The region in the $(\Phi,I)$ plane where the BS-SQUID can be successfully exploited as a tunable supercurrent diode corresponds to the area where, at constant flux, a constant-current cut-line can cross the dissipative/non-dissipative boundary  (dashed-white boundary in Fig. \ref{fig:fig5}a). When this condition is not fulfilled the device is either always superconducting or resistive. The cut lines of the color plot at selected values of the constant current are shown in Fig. \ref{fig:fig5}b: due to the skewed nature of the $I^{+,-}(\Phi_{ex}$ of the BS-SQUID, the voltage response is very far from the quasi-sinusoidal behavior of a conventional SQUID. This feature is usually conveniently exploited in feed-back loop SQUID \cite{10.1063/1.103650,MUCK2002141,133723,10.1063/1.114499,5ecc143f09064b7ca70c9bd700bb9f42,0fc1f52cc62447a3b25e1f946ca70262,Xie_2010,WANG2012127,Zhang_2013} to enhance the device transfer function $\partial V / \partial \Phi_{ex}$. From this point of view, the BS-SQUID is equivalent to a conventional additional positive feedback superconducting interferometer. The cut-lines of the color plot at selected values of $\Phi_{ex}$, shown in Fig. \ref{fig:fig5}c, on the other hand, allow us to appreciate the tunability of the BS-SQUID as a supercurrent diode. These curves, indeed, represent the current-voltage characteristics (I-V) of the device at selected values of the external flux. The net effect of sweeping the latter is equivalent to a shift in the current of device $I-V$. This mechanism can indeed be exploited to tune and reverse the polarity of the diode effect.

\begin{figure}[ht]
  \includegraphics[width=1\columnwidth]{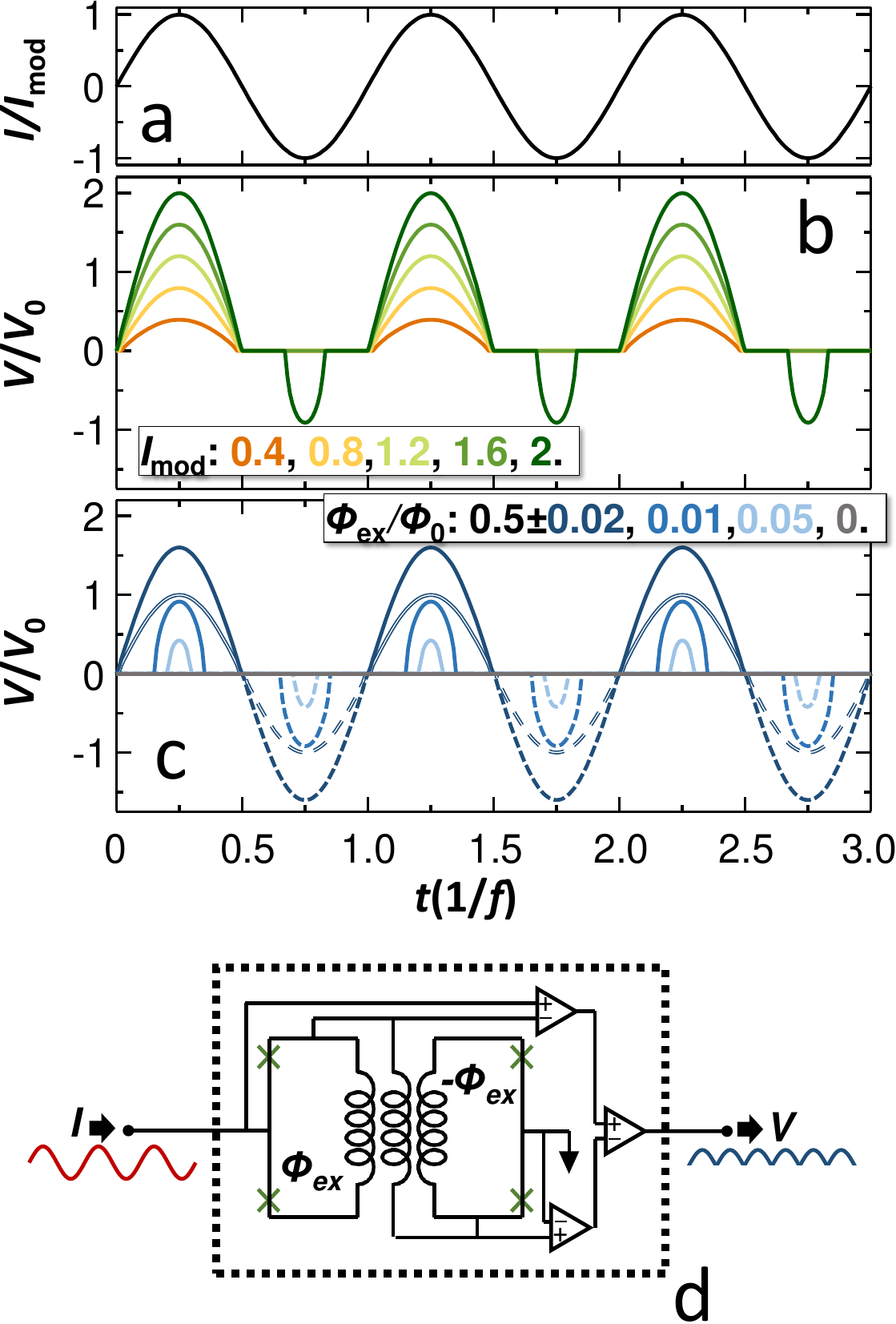}
    \caption{\textbf{Half-wave supercurrent rectification.} \textbf{a}: Normalized sinusoidal current-bias signal \textit{vs.}  time $t$ (normalized on signal  period $1/f$), applied to a BS-SQUID to calculate its time-resolved response. \textbf{b}: Normalized voltage drop $V$ of a perfectly symmetric BS-SQUID, with $\beta_l=0.0098$ and $n=36$, for selected amplitudes of the AC current bias and $\Phi{ex}=0.052$. \textbf{c}: Normalized voltage drop $V$ for $I_{mod} = 1.6 I_0$ and for selected values of $\Phi_{ex}$: solid and dashed curves correspond to positive and negative shifts from $\Phi_{ex}=0.5$, respectively. \textbf{d}: Scheme of a possible implementation of a full-wave supercurrrent rectifier based on the series of two BS-SQUID, sharing the same feedback inductance.}
  \label{fig:fig6}
\end{figure}

We can discuss a practical application of a supercurrent rectifier based on the BS-SQUID scheme. 
Specifically, we focus on rectifying a sinusoidal signal applied as a current bias to the device. 
This use case involves the detection of a radio frequency signal collected through a superconducting antenna and rectified through the BS-SQUID. A generic monochromatic current signal can be written as $I=I_{mod} sin(\frac{2 \pi t}{f})$, where $f$ is the signal frequency, $t$ is the time coordinate, and $I_{mod}$ is the peak-to-peak amplitude of the signal (see Fig. \ref{fig:fig6}a). We can model the temporal output of the BS-SQUID using Eq. \ref{eq:McCumber}. Figure \ref{fig:fig6}b displays $V(T)$ for selected values of $I_{mod}$ when $\beta_l=0.098$, $n=36$, and $\phi_{ex}=0.52 \Phi_0$. With the chosen parameters, the BS-SQUID exhibits quasi-ideal rectification ability that can efficiently rectify the signal up to $I_{mod} \lesssim 1.6 I_0$. However, beyond this threshold, a dissipative regime occurs in the \textit{reverse} half-period (e.g., dark green curve in Fig. \ref{fig:fig6}b), resulting in a reduction of the net average DC output voltage. By adjusting the flux working point, we can tune the rectification efficiency and eventually reverse the polarity of the diode.

In the figure shown as Fig. \ref{fig:fig6}c, we can see the output signal of a BS-SQUID that operates in the same conditions as the one shown in panel (b). When we move away from the ideal working point of $\Phi_{ex}\sim (0.5 \pm 0.02) \Phi_0$, the fraction of the rectified signal is reduced. Crossing the value of $0.5\Phi_0$ reverses the polarity of the diode (as seen in the dashed curves in Fig.  \ref{fig:fig6}a), allowing the rectification of the second half-period of the signal. Based on this observation, we can say that coupling two SQUIDs to the same feedback loop, as shown in Fig. \ref{fig:fig6}d, can help us realize a full-wave rectifier without increasing the inductance of the device significantly. This observation is particularly relevant because the feedback inductor and the tunnel resistance impact the device's cutoff frequency of operation due to its reactive behavior. Assuming $n=36$ and $d=1 \mu m$, we can estimate $L_S \simeq 1.6$ pH and $L_F \simeq 2$ nH. To maximize $\eta$ by assuming $\beta_l \sim 0.01$, $I_0$ needs to be $\sim6 \mu$A, which can be routinely obtained with Al/AlOx/Al JJs by setting a tunnel resistance of $\sim 200 \Omega$ during fabrication. The cutoff frequency due to the reactive component of the BS-SQUID impedance is, therefore, $f_l=R_t/L_F \sim 100$ GHz. Assuming a junction area of the order of $2 \mu$m$^2$, we can also estimate the junction capacitance $C \sim 0.2$ pF and the consequent cutoff frequency due to the capacitive impedance $f_C=1/R_tC\sim 50$ GHz, which results in the most limiting timescale with this parameter choice.

\section{Conclusions}
\label{sec:Conclusions}
In summary, we have proposed a superconducting quantum interference device (SQUID) equipped with a feedback loop that we call the boot-strap SQUID. This device is based on a magnetic flux back-action that produces a robust nonreciprocal current-voltage characteristic. By properly choosing the external magnetic flux $\Phi_{ex}$, our device can pull the supercurrent rectification coefficient to nearly unit value, resulting in a quasi-ideal supercurrent diode. Additionally, the magnetic 
flux knob can be utilized to tune and reverse the supercurrent rectification polarity. We discussed the temperature evolution of the BS-SQUID architecture and its resilience to fabrication imperfections. We also presented a possible use case as a half- or full-wave signal rectifier with a cut-off frequency of a few tens of GHz.


%% file: aknow.tex
We acknowledge the EU’s Horizon 2020 Research and Innovation Framework Programme under Grant No. 964398 (SUPERGATE), No. 101057977 (SPECTRUM), and the PNRR MUR project PE0000023-NQSTI for partial financial support.

%% file: bib.tex
%